\documentstyle[12pt,axodraw]{article}
\newcommand{\la}[1]{\label{#1}}
\newcommand{\be}{\begin{equation}}
\newcommand{\ee}{\end{equation}}
\newcommand{\ba}{\begin{eqnarray}}
\newcommand{\ea}{\end{eqnarray}}
\newcommand{\bi}{\begin{itemize}}
\newcommand{\ei}{\end{itemize}}
\newcommand{\rmi}[1]{{\mbox{\scriptsize #1}}}
\newcommand{\fig}{Fig.~}
\newcommand{\eq}{Eq.~}
\newcommand{\eqs}{Eqs.~}
\newcommand{\nr}[1]{(\ref{#1})}

\newcommand{\nn}{\nonumber \\}
\newcommand{\fr}[2]{{\frac{#1}{#2}}}
\newcommand{\msbar}{\overline{\mbox{\rm MS}}}

\newcommand{\bmu}{\bar{\mu}}
\newcommand{\im}{\mathop{\mbox{Im}}}
\renewcommand{\vec}[1]{{\bf #1}}

\def\lsi{\raise0.3ex\hbox{$<$\kern-0.75em\raise-1.1ex\hbox{$\sim$}}}
\def\gsi{\raise0.3ex\hbox{$>$\kern-0.75em\raise-1.1ex\hbox{$\sim$}}}

\newcommand{\gsim}{\mathop{\gsi}}

\begin{document}

\begin{titlepage}
\begin{flushright}
CERN-TH/99-62\\
hep-ph/9903513 
\end{flushright}
\begin{centering}
\vfill

{\bf THE RENORMALIZED GAUGE COUPLING AND\\ 
NON-PERTURBATIVE TESTS OF DIMENSIONAL REDUCTION}
\vspace{0.8cm}

M. Laine
\footnote{mikko.laine@cern.ch}

\vspace{0.3cm}
{\em 
Theory Division, CERN, CH-1211 Geneva 23,
Switzerland\\}
\vspace{0.3cm}
{\em 
Dept.\ of Physics,
P.O.Box 9, FIN-00014 Univ.\ of Helsinki, Finland\\}

\vspace{0.7cm}
{\bf Abstract}

\end{centering}

\vspace{0.3cm}\noindent
In 4d lattice simulations of Standard Model like theories, the
renormalized gauge coupling in the broken phase can be determined from 
the prefactor of the Yukawa term in the static potential. We compute
the same quantity in terms of the conventional $\msbar$ scheme gauge 
coupling. The result allows for a further non-perturbative test of 
finite temperature dimensional reduction, by a comparison of the 
critical temperatures for the electroweak phase transition as
obtained with 4d lattice simulations and 
with 3d effective theory simulations. 

\vfill

\noindent
CERN-TH/99-62\\
June 1999

\vfill

\end{titlepage}

\section{Introduction}

Consider finite temperature physics at temperatures $\pi T\gg m$, 
where $m$ stands for the mass scales of the problem. Such a case is
for instance the electroweak phase transition at $T_c \sim m_H/g$, 
with a Higgs vev $v(T_c) \sim T_c$~\cite{rs}. 
One can then construct a three-dimensional (3d) effective 
theory for the thermodynamics of the system with the method 
of dimensional reduction~\cite{dr}--\cite{jp}. The construction
is purely perturbative, while the non-perturbative IR-problems
of finite temperature field theory are contained in the effective theory. 
Often, even further degrees of freedom can be integrated out
within the 3d theory~\cite{dr}--\cite{jp}.

For the electroweak sector of the Standard Model
(as well as for many extensions thereof~\cite{generic}), 
the final effective action resulting from the procedure 
described above is of the very simple form
\be
S_\rmi{eff} = \int d^3 x \biggl[
{1\over4} F_{ij}^aF_{ij}^a+{1\over4} B_{ij}B_{ij}+
(D_i\phi)^\dagger D_i\phi+m_3^2\phi^\dagger\phi+
\lambda_3(\phi^\dagger\phi)^2 \biggr],
\label{contaction}
\ee
where $F_{ij}^a$ and $B_{ij}$ are the SU(2) and U(1) field
strength tensors. The dynamics of this theory depends on 
the SU(2) and U(1) gauge couplings $g_3^2,g_3'^2$, 
as well as on the Higgs sector parameters $m_3^2,\lambda_3$ 
(or, more precisely, on the three dimensionless ratios that 
can be formed thereof). The practical content of dimensional
reduction is to compute these parameters as
a perturbative expansion in the underlying physical
4d parameters and the temperature;  explicit derivations 
have been carried out in~\cite{generic,mssm}. 
For instance, the expression for $m_3^2$ 
is parametrically of the form 
\be
m_3^2 = m^2(\bmu)\Bigl[ 1 + {\cal O}(g^2)\Bigr] + 
g^2(\bmu) T^2 \Bigl[ 1 + {\cal O}(g) + {\cal O}(g^2)\Bigr], 
\la{m32}
\ee
where the parameters appearing 
on the right hand side
are those of the corresponding
4d theory in the $\msbar$ scheme (see below), 
with $\bmu$ the $\msbar$ scale parameter, and all the 
leading ${\cal O}(g,g^2)$ corrections shown have been explicitly computed
for several theories.  

Of course, the final 3d theory obtained with dimensional reduction, 
such as the one in \eq\nr{contaction}, 
is ``only'' an effective theory, 
and it is not arbitrarily precise. Analytically, the accuracy 
of dimensional reduction can be estimated by considering
the set of one-particle-irreducible Green's functions for the
degrees of freedom contained in \eq\nr{contaction},  
and comparing the effects of the 
higher-dimensional operators left out from the effective theory, 
with those arising within the effective theory. The conclusion is 
that the relative error remaining in the non-vanishing P- and CP-even
static bosonic Green's functions 
with ``soft'' external momenta
is highly suppressed, ${\cal O}(g^3)$, already 
when only super-renormalizable operators are kept in the 
effective theory~\cite{generic}
(as is the case with \eq\nr{contaction}). 
Numerically, the error
was estimated to be at the percent level for the Standard
Model (the largest corrections arising from 
the top loops, and from the Matsubara zero modes of the 
temporal components of the gauge fields)~\cite{generic}.

However, many of the quantities addressed with the effective theory
are purely non-perturbative. Thus, it is 
strictly speaking impossible
to estimate the numerical accuracy analytically. Even though it is 
clear that the non-perturbative effects can only arise 
from the degrees of freedom contained in the
3d theory, one may still ask what the relative accuracy 
obtained is for such quantities. While in the electroweak case
this is not feasible 
within the full Standard Model due to chiral fermions, the 
question can at least be addressed within the SU(2)+Higgs model
with 4d finite temperature lattice simulations
(the U(1) subgroup is not too essential for these
considerations~\cite{nonpert}).

The 4d simulations relevant for studying the 
accuracy of the theory in \eq\nr{contaction},
have been carried out with a gauge coupling which 
is $\sim$50\% larger than the physical one,
$g^2 \approx 0.585$~\cite{bdsim1,bdsim2}. This
should increase the possible discrepancies. However, 
the system still has multiple length scales ($\sim \pi T, gT, g^2T$), 
making the extrapolation of the 
4d results to the continuum limit extremely demanding (but 
unavoidable if one wants to compare with the 3d theory).
Consequently, the extrapolation has been 
carried out in the full range of relevant 
Higgs masses only very recently~\cite{bdsim2}.

The problem we study here is a systematic 
comparison of the 4d lattice results in~\cite{bdsim1,bdsim2}
with the 3d effective theory results in~\cite{nonpert,isthere}%
\footnote{A similar comparison was carried out in~\cite{4d3d}, 
but at that time a 4d continuum extrapolation existed
only for a single Higgs mass~\cite{bdsim1}, and the
relation to be computed in Sec.~3 was not available.}.
In order to make such a comparison, we will need to perform
one further (well convergent) zero temperature perturbative
computation, to which we now turn. 

\section{Formulation of the problem}

Dropping chiral fermions and 
the U(1) subgroup from the standard electroweak theory, 
we consider the 4d SU(2)+Higgs model in this paper:
\be
S = \int d^4 x \biggl[ 
\fr14 F^a_{\mu\nu} F^a_{\mu\nu} + 
(D_\mu \phi)^\dagger (D_\mu \phi) + 
m^2 \phi^\dagger \phi + 
\lambda (\phi^\dagger \phi)^2\biggr], \la{action}
\ee
where $D_\mu = \partial_\mu + i g T^a A^a_\mu$.
The theory in \eq\nr{action}  
has three parameters: the scalar 
mass parameter $m^2$, the scalar self-coupling $\lambda$ and the gauge 
coupling $g^2$. To fix a particular physical 
theory, one has to choose a regularization 
scheme and then give the values of the renormalized couplings in this
scheme in terms of some physical quantities. As a regularization
scheme we choose $\msbar$ as is convenient in continuum computations, 
and in particular in dimensional reduction~\cite{generic}. 
The question we address then is,
what are the values of the $\msbar$ parameters if a set of 
physical observables is determined, either by 
experiment or by 4d lattice simulations?  
 
In the broken phase of the theory, two of the $\msbar$ couplings
can then be fixed in terms of the physical masses 
of the Higgs and the W boson, $m_H$ and $m_W$, respectively: 
\be
m^2(\bmu) = -\fr12 m_H^2 + \delta m^2(\bmu), 
\quad \lambda(\bmu) = 
\frac{g^2(\bmu)}{8}\frac{m_H^2}{m_W^2}+\delta \lambda(\bmu), \la{msrel}
\ee
where the 1-loop
corrections $\delta m^2(\bmu), \delta \lambda(\bmu)$ are easily computable 
(we employ here the formulas given in~\cite{generic}). However, 
the value of the gauge coupling $g^2(\bmu)$ 
cannot be fixed in terms of the masses. 
In the physical case with fermions, $g^2(\bmu)$ can be fixed, e.g., 
through the muon lifetime; for an explicit expression
see \eq(183) in~\cite{generic}. However, 
this is not available in the theory of \eq\nr{action}. 

We thus need another physical observable sensitive to $g^2$. 
Moreover, since we are ultimately
interested in lattice studies of the theory in \eq\nr{action},
this observable should be measurable in Monte Carlo simulations. 
A suitable choice is 
the ``renormalized gauge coupling $g_R^2$''~\cite{bdsim0}. 
The value of $g_R^2$ is obtained 
from the static potential $V(r)$: 
from a large rectangular Wilson loop $W(r,t)$ of
size $r\times t$ (in Euclidian space), one determines 
\be
V(r) = \lim_{t\to\infty} -\frac{1}{t} \ln W(r,t). \la{potdef}
\ee
At leading order, the potential thus defined is 
\be
V_\rmi{tree} (r) = 
- g^2 C_F \int \frac{d^3 p}{(2 \pi)^3} \frac{e^{i\vec{p}\cdot\vec{r}}-1}
{p^2 + m_W^2} = -g^2 C_F \frac{e^{-m_W r}}{4 \pi r} + 
V_\rmi{tree} (\infty),  \la{Vtree}
\ee
where $C_F = (N^2-1)/(2 N) = 3/4$ and
$V_\rmi{tree} (\infty)$ is a regularization dependent constant. 
Let us now define
\be
g_R^2(r) = \frac{1}{C_F}\frac{\frac{d}{dr} \Bigl[ -V(r)\Bigr]}
{\frac{d}{dr} \int \frac{d^3 p}{(2 \pi)^3} \frac{e^{i\vec{p}\cdot\vec{r}}}
{p^2 + M^2}}, \la{gRdef}
\ee
where $M$ is some physical mass parameter, 
chosen in~\cite{bdsim0,bdsim1} to be the measured 
coefficient of the exponential falloff of $V(r)$. 
At leading order, $M$ coincides with $m_W$. In~\cite{bdsim1,bdsim2}, 
the distance used in $g_R^2(r)$ was further fixed to be $r=M^{-1}$. 
Note that the choice of $M$ in the dominator of \eq\nr{gRdef}
has a 1st order effect on $g_R^2$, $\delta g_R^2 \propto
g^2 (\delta M/M)$, while the choice of $r$ 
has only a 2nd order effect, $\delta g_R^2 \propto
g^2 (\delta r/r)^2$ (see below).

The expression in \eq\nr{gRdef} is the continuum version of 
the definition given in~\cite{bdsim0}: we do not discuss
the finite lattice spacing artifacts, but assume that 
the extrapolation to the continuum limit performed 
in~\cite{bdsim1} is reliable. 
To leading order, the result of the definition in \eq\nr{gRdef}
is just $g^2(\bmu)$, but $g_R^2(M^{-1})$ 
is defined also fully non-perturbatively.

Our objective is now to find the relation of 
$g^2(\bmu)$ and $g_R^2(M^{-1})$ at 1-loop level
(for clarity, we often denote $g^2(\bmu)=g^2_\rmi{$\msbar$}(\bmu)$). 
Both quantities are finite and not sensitive to the 
ultraviolet regularization, so that the computation can
be carried out in the continuum. In this way, we have found the 
$\msbar$ parameters to which the 4d simulations correspond (Sec.~3). 
The $\msbar$ parameters are, in turn, the input in 
dimensional reduction and the construction of 3d effective
theories, so that with the relation found, 
we can compare the results of direct 4d finite temperature
lattice simulations
with those of 3d effective theory simulations (Sec.~4).

Finally, it should be mentioned that 
there is also another approach available
for determining the $\msbar$ scheme parameters to which
the 4d simulations correspond. 
Indeed, instead of going via a set of 
physical observables, one could directly relate the bare 
lattice parameters at a finite but small lattice spacing $a$,
and the renormalized $\msbar$ parameters, as first done for QCD
in~\cite{hh}. In the continuum limit
the end results of the two approaches are in principle
equivalent up to higher order effects. 
However, the convergence properties
can be somewhat different: relating the regularization
schemes is a computation only sensitive to the ultraviolet, 
while the computation we carry out is only sensitive to 
the infrared. The reason why we choose our approach is that the 
extrapolation to the continuum limit has in~\cite{bdsim1,bdsim2}
been carried out directly for the physical observables $m_H,m_W,g_R^2$, 
so that this way we get rid of any 
reference to a finite lattice spacing, 
as is necessary for a comparison of 4d and 3d lattice results. 
The 1-loop computation we perform is very well convergent despite the 
infrared sensitivity, since we are in the broken phase of the theory. 
It should be noted, though, that for the academic case of
very small Higgs masses close to the Coleman-Weinberg limit
$m_H\approx 10$ GeV, the ``strict loop expansion''
formulas~\cite{generic} we employ here for relating 
$m_H,m_W$ to $m^2(\bmu),\lambda(\bmu)$ break down
(see Sec.~4)\footnote{There is no conceptual problem 
in finding relations which are accurate also for very 
small Higgs masses, but this exercise is beyond
the scope of the present paper.}.

\section{Computation of $g_R^2$}
\la{sec:comp}

{\bf Preliminaries.} 
We carry out the computation in  
a general 't Hooft $R_\xi$-gauge with the gauge parameter $\xi$
and, for the moment, a general spacetime dimension $d$.
We denote the different 
structure constants appearing as
\ba
& & (T^aT^a)_{ij} = C_F \delta_{ij}, \quad
f^{abc} f^{abd} = C_A \delta^{cd}, \quad 
T^aT^bT^aT^b = C_F^2-\fr12 C_FC_A,
\ea
where  
$C_F = (N^2-1)/(2 N), C_A=N$.
The actual computation is carried out for $N=2$, but
we keep the symbols $C_F,C_A$ in some of the formulas below, 
to differentiate between two classes 
of contributions (Abelian and non-Abelian; see 
the Appendix).

The object we compute is the expectation value of a
rectangular Wilson loop in the limit $t\to\infty$, \eq\nr{potdef}. 
It can be seen that the horizontal parts of the path ($t=$ const.)\ 
do not contribute in this limit~\cite{wf,de}. Thus, denoting
\be
U(-\vec{r}/2,t)= 
{\cal P} \exp\Bigl[ig \int_{-t/2}^{t/2} dt' T^a A^a_0(-\vec{r}/2,t')\Bigr], 
\la{Udef}
\ee
where the path-ordered exponential is defined through
\be
{\cal P} \exp\Bigl[\int_{-T}^{T} dt' M(t')\Bigr] = 
1+\int_{-T}^{T} dt' M(t')
+\int_{-T}^{T} dt' dt'' 
\theta (t'-t'')  M(t'') M(t') + \ldots, \la{Pexp}
\ee
the expectation value to be computed is 
\be
\lim_{t\to\infty} W(r,t) =
\lim_{t\to\infty} \langle U(-\vec{r}/2,t) U^\dagger(\vec{r}/2,t) \rangle. 
\la{Wrt}
\ee

The actual computation proceeds in a straightforward way, by inserting 
the expansion in \eq\nr{Pexp} into \eq\nr{Wrt}. 
The $\theta$-functions can be written as usual as 
\be
\theta(t) = 
\int \frac{d\omega}{2\pi} \frac{e^{i\omega t}}{i \omega + \epsilon^+}.
\ee
The $t'$-integral in \eq\nr{Udef} makes the sources 
static in the limit $t\to\infty$, and in momentum ($\omega$) 
space, the graphs then produce a term linear in $t$ via
\be
2\pi\delta(0) \to t.
\ee
Through \eq\nr{potdef}, the coefficient of the linear term gives $-V(r)$. 
This computation is, of course, a direct generalization of the 
one in~\cite{wf} for QCD (for 2-loop results in QCD, see~\cite{de}). 
The graphs needed are shown in~\fig\ref{graphs}, and the results
for these individual graphs are given in the Appendix.

\begin{figure}[t]


\hspace*{1cm}%
\begin{picture}(350,50)(0,0)
\SetWidth{1.5}
\SetScale{1.15}

\Line(10,0)(10,50)
\Line(40,0)(40,50)
\Photon(10,15)(40,15){1.5}{6}
\Photon(10,35)(40,35){1.5}{6}

\Line(60,0)(60,50)
\Line(90,0)(90,50)
\Photon(60,15)(90,35){1.5}{7}
\Photon(60,35)(90,15){1.5}{7}

\Line(110,0)(110,50)
\Line(140,0)(140,50)
\Photon(110,10)(140,10){1.5}{6}
\PhotonArc(110,30)(10,90,270){1.5}{6}

\Line(160,0)(160,50)
\Line(190,0)(190,50)
\Photon(160,25)(190,25){1.5}{6}
\PhotonArc(160,25)(10,90,270){1.5}{6}

\Line(210,0)(210,50)
\Line(240,0)(240,50)
\Photon(225,25)(240,25){1.5}{3}
\Photon(210,35)(225,25){1.5}{4}
\Photon(210,15)(225,25){1.5}{4}

\Line(260,0)(260,50)
\Line(290,0)(290,50)
\Photon(260,25)(290,25){1.5}{6}
\GCirc(275,25){6}{0.5} 

\Line(310,0)(310,50)
\Line(340,0)(340,50)
\Photon(310,25)(340,25){1.5}{6}
\GCirc(310,25){3}{0.0} 

\Text(23,-15)[l]{(a)}
\Text(80.5,-15)[l]{(b)}
\Text(138,-15)[l]{(c)}
\Text(195.5,-15)[l]{(d)}
\Text(253,-15)[l]{(e)}
\Text(310.5,-15)[l]{(f)}
\Text(368,-15)[l]{(g)}

\end{picture}

\vspace*{1.0cm}

\caption[a]{
The graphs contributing to the static potential at 1-loop level.
``Mirror'' configurations are not shown. Graph (g)
is a counterterm contribution.}
\la{graphs} 
\end{figure}
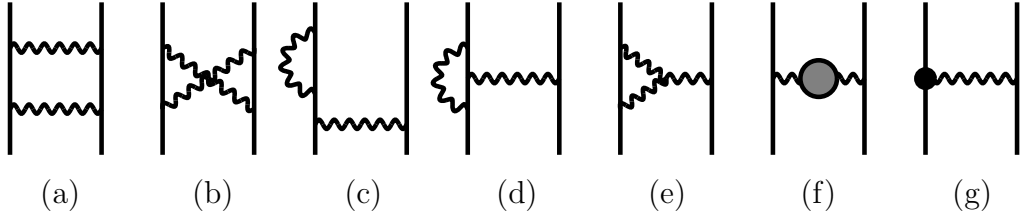

\vspace*{0.3cm}

{\bf The loop contributions.} 
The final result
for the computation of the graphs in \fig\ref{graphs}
can be expressed in terms of the functions
\ba
A_0 (m^2) & = & 
\int \frac{d^dp}{(2\pi)^d} \frac{1}{p^2+m^2}, \\
B_0 (k^2;m_1^2,m_2^2) & = &  
\int \frac{d^dp}{(2\pi)^d} \frac{1}{[p^2+m_1^2][(p+k)^2 + m_2^2]}.
\ea
Summing all the graphs together, we obtain (for $N=2$) 
\ba
& & -V_\rmi{1-loop}(r) = g^4 C_F C_A \int \frac{d^{d-1} p}
{(2\pi)^{d-1}}
\frac{e^{i \vec{p}\cdot\vec{r}}}{p^2+m_W^2}\biggl\{ \nn 
& & \hspace*{1cm} 
2B_0(0;m_W^2,m_W^2) +  (p^2+m_W^2) 
\frac{d}{d m_W^2} B_0(p^2;m_W^2,m_W^2) \nn
& &  \hspace*{1cm}+ \frac{1}{8(d-1)(p^2+m_W^2)} \biggl[
B_0(p^2;m_W^2,m_W^2) \Bigl(
 3(4d-3){p^2}+ 12(3-2d){m_W^2}\Bigr) \nn
& &  \hspace*{1cm} - 
B_0(p^2;m_W^2,m_H^2) \frac{1}{p^2} 
\Bigl( {p^4} +(6-4d){p^2}{m_W^2}+ 
         2p^2 {m_H^2}
         +({m_W^2} - {m_H^2})^2  \Bigr) \nn
& &  \hspace*{1cm} + 
A_0(m_W^2)\frac{1}{p^2 m_H^2} 
\Bigl(  6{{(d-1) }^2}{p^2}{m_W^2}- 
         p^2  m_H^2  ( 29 - 32d + 8{d^2})
         + {m_H^4} - {m_H^2} m_W^2   \Bigr) \nn
& &  \hspace*{1cm} + 
A_0(m_H^2) \frac{1}{p^2} \Bigl(
  (2 d -1) p^2 + m_W^2 - {m_H^2}  \Bigr)\biggl]
  +\, {\rm counterterms} \biggr\}.
\la{res1}
\ea
This result is, of course, independent of the gauge parameter $\xi$, 
even though the single contributions given 
in the Appendix do depend on it.

Specializing then to $d=4-2\epsilon$, 
$A_0, B_0$ have their standard forms
\ba
A_0 (m^2) \!\! & = & \!\! -\frac{\mu^{-2\epsilon}}{(4\pi)^2} m^2
\left( \frac{1}{\epsilon} + \ln\frac{\bmu^2}{m^2}+1 \right), \\
B_0 (k^2;m_1^2,m_2^2) \!\! & = &  \!\! \frac{\mu^{-2\epsilon}}{(4\pi)^2}
\biggl[
\frac{1}{\epsilon} + \ln\frac{\bmu^2}{m_1 m_2}+1 -
\frac{m_1^2+m_2^2}{m_1^2-m_2^2} \ln \frac{m_1}{m_2} + 
F_E(k^2; m_1^2,m_2^2)
\biggr], \mbox{\hspace*{0.8cm}}
\ea
where
\ba
& & F_E(k^2; m_1^2,m_2^2) = 
1+\frac{m_1^2+m_2^2}{m_1^2-m_2^2} \ln \frac{m_1}{m_2}+ 
\frac{m_1^2-m_2^2}{k^2} \ln \frac{m_1}{m_2} \nn
& & \hspace*{1cm} +  
\frac{1}{k^2} 
\sqrt{(m_1+m_2)^2+k^2}
\sqrt{(m_1-m_2)^2+k^2} \ln
\frac{1-\sqrt{\frac{(m_1-m_2)^2+k^2}{(m_1+m_2)^2+k^2}}}
{1+\sqrt{\frac{(m_1-m_2)^2+k^2}{(m_1+m_2)^2+k^2}}}.
\la{FE}
\ea
Including also the counterterms which cancel the
$1/\epsilon$ divergences, we obtain
\ba
-V_\rmi{1-loop}(r) & = & \frac{g^4}{16\pi^2} 
C_F C_A \int \frac{d^{3} p}
{(2\pi)^{3}}
\frac{e^{i \vec{p}\cdot\vec{r}}}{p^2+m_W^2}\biggl\{ \nn 
& &  \frac{p^2 + m_W^2}{p^2} \frac{2}{\sqrt{1+\frac{4 m_W^2}{p^2}}}
\ln\frac{1-(1+\frac{4 m_W^2}{p^2})^{-1/2}}
{1+(1+\frac{4 m_W^2}{p^2})^{-1/2}} \nn 
& + & \frac{1}{p^2+m_W^2} \biggl[
 \frac{1}{24 h^2} \Bigl( 86{h^2}{p^2}
       -9  ( 6 - 3h^2 + {h^4} )  
         {m_W^2}  \Bigr)  
  \ln\frac{\bmu^2}{m_W^2} \nn
& + & \fr18 (13{p^2}  -20{m_W^2} ) F_E(p^2;m_W^2,m_W^2) \nn
& - & \frac{1}{24} \Bigl( 
      (h^2-1)^2 \frac{m_W^4}{p^2}+ {p^2} + 
        2(h^2-5){m_W^2} \Bigr)
        F_E(p^2;m_W^2,m_H^2) \nn
& + & \frac{h^2 \ln h}{12(h^2-1)} \Bigl( {p^2}+
   (9{h^2}-17) {m_W^2}  \Bigr)  \nn
& + & \frac{1}{72 h^2} 
   \Bigl( {h^2}{p^2} + 
      3 ( -18 + {h^2} - 11{h^4}) 
       {m_W^2}  \Bigr)\biggr]\biggr\},
\la{res2}
\ea
where $h=m_H/m_W$.

\vspace*{0.3cm}

{\bf The Fourier transformation.}
It remains to evaluate \eq\nr{gRdef}, i.e., to take the derivative with 
respect to $r$ and to perform the final integral with respect to $\vec{p}$. 
Let us note here that since the difference
between $M$ and $m_W^\rmi{tree}$ is of relative order $g^2$ and the 
scale dependence of $g^2_\rmi{$\msbar$}(\bmu)$ is already of 
order $g^4$, we can replace $M^{-1}$ by $(m_W^\rmi{tree})^{-1}$ in 
the argument of $g_R^2(r)$ in the present 1-loop computation.
For the time being we denote the tree-level
value of the W mass by $m_W=m_W^\rmi{tree}$. Within 1-loop
corrections, the difference between $m_W^\rmi{tree}$ and
$m_W^\rmi{1-loop}$ is naturally of higher order. 

\begin{figure}[t]

\vspace*{-0.5cm}

\begin{center}
\begin{picture}(100,100)(0,0)
\SetScale{1.0}

\SetWidth{0.5}
\Line(-50,0)(150,0)
\Line(50,-20)(50,100)
\SetWidth{1.5}
\GCirc(50,25){2}{0.0} 

\SetWidth{2.0}
\Line(50,50)(50,100)
\SetWidth{1.0}

\CArc(50,25)(6,0,180)
\ArrowArc(50,25)(6,180,360)
\ArrowArc(50,50)(6,180,360)
\Line(56,50)(56,100)
\Line(44,50)(44,100)

\Text(62,95)[l]{Im($p_z$)}
\Text(120,10)[l]{Re($p_z$)}

\end{picture}
\end{center}


\caption[a]{
The integration contour in the $p_z$-plane.}
\la{plane} 
\end{figure}
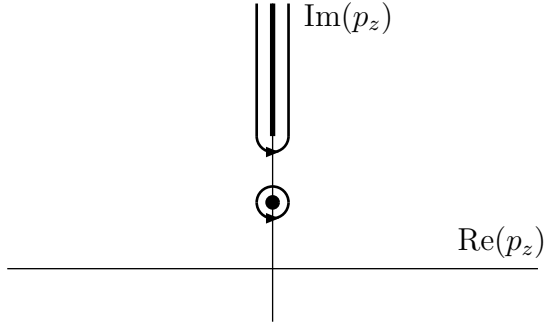

The final $\vec{p}$-integral to be evaluated is convergent 
and can be done numerically by brute force, but it can 
be put in a more illuminating and rapidly convergent
form by doing a part of it by contours. 
We write $\vec{p}=(p_z,\vec{p}_\perp)$, such that 
$\vec{p}\cdot\vec{r} = p_z r$, and close the $p_z$-integral in 
the upper half plane. 
The analytic structure of the $p_z$-integrand is that there
is a (1st or 2nd order) pole at  $p_z = +i ({p}_\perp^2 + m_W^2)$, 
and cuts along the imaginary axis starting
at some $p_z= i p_z^0$, see \fig\ref{plane}.

\vspace*{0.3cm}

{\bf The pole contribution.} The pole contribution is of the form
\be
\left. \frac{d}{dr} 
\int \frac{d^3 p}{(2 \pi )^3}\frac{e^{i p_z r}}{(p^2 + m_W^2)^2} 
{\cal F}(p^2+m_W^2)\right|_{r=m_W^{-1}} = 
-\frac{1}{8 \pi} e^{-1} \Bigl[ 
{\cal F}(0) + 4 {\cal F}'(0) m_W^2 \Bigr]. \la{eq:pole}
\ee
For the expression in the square brackets in \eq\nr{res2} 
(the term on the 2nd row in \eq\nr{res2} does not contribute),
\ba
{\cal F}(0) & = & m_W^2 \biggl[
  \left( -{\frac{59}{24}} - {\frac{9}{{4h^2}}} - \frac{3}{8} {h^2}
     \right) \ln\frac{\bmu^2}{m_W^2} \nn 
& -  &  
    \frac{h}{12} 
      \left( 12 - 4{h^2} + {h^4} \right) \sqrt{4-h^2 } 
\arctan {\sqrt{{\frac{2 - h}{2 + h}}}} \nn
& - &  \frac{1}{24} {h^4}({h^2-6}) \ln h + 
\frac{1}{72 h^2} \left(
 -54+ {h^2} ( -556 + 99 {\sqrt{3}} \pi)
 - 45 {h^4} + 3 {h^6} \right)\biggr] \nn
 & = & -\fr12 \biggl(\frac{g^2}{(4\pi)^2}
\biggr)^{-1}\delta m_W^{2{\rm(1-loop)}}, \la{F0} \\
{\cal F}'(0) & = & 
  {\frac{43}{12}} \ln\frac{\bmu^2}{m_W^2} \nn 
& + &  \frac{h}{12\sqrt{4-h^2}}
       \left( -36 + 32{h^2} - 13{h^4} + 2{h^6} \right) 
          \arctan {\sqrt{{\frac{2 - h}{2 + h}}}} \nn
& + & \frac{1}{24}
      \left( 12 - 18{h^2} + 9{h^4} - 2{h^6} \right) \ln h 
- \frac{1}{72} (26 +18{h^2} - 6{h^4} - 27{\sqrt{3}}\pi). \la{Fp0}
\ea
In \eq\nr{F0}, $m_W^2+\delta m_W^{2{\rm(1-loop)}}$ denotes
the 1-loop W pole mass squared.

\vspace*{0.3cm}

{\bf The cut contribution.} The cut contribution is of the form
\be
\int_{-\infty}^{\infty} dp_z {\cal F}(p_z) \to 
\int_{p_z^0}^{\infty} i dp_z \Bigl[ {\cal F}(i p_z+ \epsilon) -    
{\cal F}(i p_z- \epsilon) \Bigr]
= - 2 \int_{p_z^0}^{\infty} dp_z \im 
{\cal F}(i p_z+ \epsilon).
\ee
Non-vanishing imaginary parts
arise from the function $F_E(p^2;m_1^2,m_2^2)$
in \eq\nr{FE}, and from the first logarithm in \eq\nr{res2}: 
\ba
& & 
\left.\im F_E (p^2 ;m_1^2,m_2^2)
\right|_{p_z \to i p_z+\epsilon} =   
-\theta(p_z^2 - p_\perp^2-(m_1+m_2)^2)  \nn 
& & \hspace*{1cm} \times
\frac{\pi}{p_z^2-p_\perp^2} \sqrt{p_z^2-p_\perp^2-(m_1+m_2)^2} 
\sqrt{p_z^2-p_\perp^2-(m_1-m_2)^2}, \\
& & 
\im \left.\frac{2}{p^2 \sqrt{1+\frac{4 m_W^2}{p^2}}}
\ln\frac{1-(1+\frac{4 m_W^2}{p^2})^{-1/2}}
{1+(1+\frac{4 m_W^2}{p^2})^{-1/2}} 
\right|_{p_z \to i p_z + \epsilon} \nn 
& & \hspace*{1cm} = 
\theta (p_z^2-p_\perp^2 - 4 m_W^2 ) \frac{2\pi}{p_z^2-p_\perp^2}
\frac{1}{\sqrt{1-\frac{4 m_W^2}{p_z^2 - p_\perp^2}}}.
\ea
The expression in the curly brackets in \eq\nr{res2} 
then contributes as follows: from the
first term and from $F_E(p^2;m_W^2,m_W^2)$, one gets a 
constant factor which, by a change of variables, can be written  
in the following rapidly convergent form:
\ba
& & f_0 = \int_2^{\infty}\! dy \int_0^{\infty}\! dz e^{-(y + z)}
y\left( y + z \right) \nn
& & \hspace*{1cm} \times \frac{144 + 336yz + 168{z^2} + 116{y^2}{z^2} + 
       116y{z^3} + 29{z^4}} 
{4 \left( 3 + 2yz + {z^2} \right)^2
 \sqrt{z\left( 2y + z \right) 
         \left( 4 + 2yz + {z^2} \right)}} = 2.156946. \la{f0} 
\ea
{}From $F_E(p^2;m_W^2,m_H^2)$, on the other hand, there
is a contribution which depends on $h$; 
it can be similarly written, e.g., in the form 
\ba
f(h) \!\! & = & \!\! -\int_{h+1}^{\infty}\! dy \int_0^{\infty}\! dz 
e^{-(y + z)} \frac{y\left( y + z \right) \sqrt{z\left( 2y + z \right) 
           \left( 4h + 2yz + {z^2} \right) }}
{12 \left[(h+1)^2 + 2yz + {z^2}\right]^2
     \left[h(h+2) + 2yz + {z^2}\right]^2}  \nn
& & \times 
\left( z^4 + 4 y z^3 + 4 y^2 z^2 + 4 (h+3) z^2+
8 (h+3) y z +12 (h+1)^2 \right). \la{fh}
\ea
The numerical values of $f(h)$ are given in 
Table~\ref{table:gR2} and \fig\ref{fig:gR2}.

\vspace*{0.3cm}

{\bf The final result.}
We are now in a position to collect all the terms together, 
to evaluate \eq\nr{gRdef}. The denominator there gives, 
to 1st order in $M-m_W\propto g^2$,
\be
\left. \frac{d}{dr} \frac{1}{4 \pi r} 
\exp \Bigl(
-M r 
\Bigr) \right|_{r= m_W^{-1}} = 
-\frac{1}{2\pi} e^{-1} m_W^2
\biggl(1 + \fr12 \frac{m_W-M}{m_W}\biggr).
\la{denom}
\ee
The order $(m_W-M)$ term here 
combines with ${\cal F}(0)$ in \eqs\nr{eq:pole}, \nr{F0}, 
to replace $m_W$ with $m_W^\rmi{1-loop}$. 
The final result left is then
\ba
g_R^2(M^{-1}) & = & 
g^2_\rmi{$\msbar$}(\bmu)
\biggl(1 + \fr12 \frac{M-m_W}{m_W} \biggr) +  
\frac{g^4}{(4 \pi)^2} \Bigl[
2 {\cal F}'(0)   
-e \Bigl(f_0 + f(h) \Bigr)
\Bigr] \nn
& \equiv & g^2_\rmi{$\msbar$}(m_W)
\biggl(1 + \fr12 \frac{M-m_W}{m_W} \biggr) +  
\frac{g^4_\rmi{$\msbar$}(m_W)}{(4 \pi)^2}
C(h), \la{Ch}
\ea
where ${\cal F}'(0)$, $f_0$, $f(h)$ are from 
\eqs\nr{Fp0}, \nr{f0}, \nr{fh}, 
and we have now restored the meaning of
$m_W$ as the physical (1-loop) pole mass.
Going from the first to the second row in \eq\nr{Ch},
we have merged the first term of ${\cal F}'(0)$
in \eq\nr{Fp0} with $g^2_\rmi{$\msbar$}(\bmu)$, 
which produces $g^2_\rmi{$\msbar$}(m_W)$ up to 
corrections which are of higher order than the 
present computation, and we have then chosen 
to use the same scale also in all the terms
of order $g^4_\rmi{$\msbar$}$, again by ignoring 
terms which are of higher order. After this 
replacement, only the latter two rows in \eq\nr{Fp0}
contribute on the latter row in \eq\nr{Ch}:
\be
C(h) = \left. 2 {\cal F}'(0)\right|_{\bmu=m_W} - 
e \Bigl(f_0 + f(h) \Bigr).
\ee 
The numerical values of $C(h)$
are shown in Table~\ref{table:gR2} and \fig\ref{fig:gR2}.
While $f(h)$ diverges logarithmically at small $h$, 
$C(h)$ is finite. 

\begin{table}[t]
\centering
\begin{tabular}{lll}
\hline
$h$ & $f(h)$  & $C(h)$ \\ \hline
   0.2  & -0.249947  & -3.78735 \\
   0.4  & -0.118355  & -3.69862 \\
   0.6  & -0.0660750  & -3.62889 \\
   0.8  & -0.0399324  & -3.57204 \\
   1.0  & -0.0253292  & -3.52459 \\
   1.2  & -0.0166083  & -3.48430 \\
   1.4  & -0.0111594  & -3.44959 \\
   1.6  & -0.00764098  & -3.41935 \\
   1.8  & -0.00531146  & -3.39272 \\
   2.0  & -0.00373821  & -3.36907 \\ \hline
\end{tabular}
\caption[a]{\protect The values of $f(h)$ and 
$C(h)$,  \eqs\nr{fh}, \nr{Ch}, at selected values of $h$.}
\la{table:gR2}
\end{table}

\begin{figure}[t]
\centerline{\epsfxsize=8cm\epsfbox{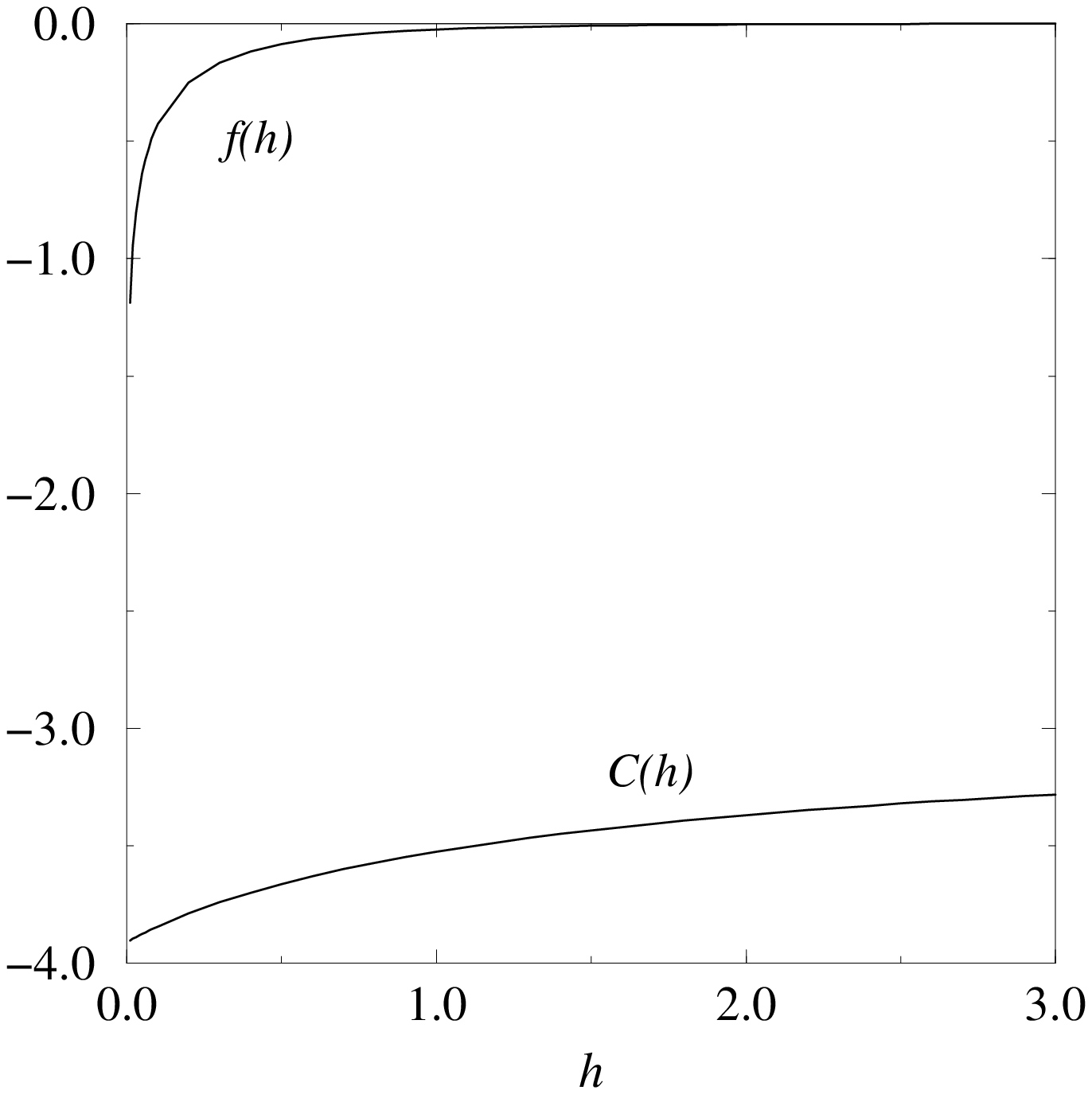}}
\caption[a]{The functions $f(h)$ and $C(h)$ from 
\eqs\nr{fh}, \nr{Ch}.}  
\la{fig:gR2}
\end{figure}

The fact that $C(h)$ is negative, means that 
$g^2_\rmi{$\msbar$}(m_W)$ tends to be 
larger than \linebreak
$g_R^2(M^{-1})$. 
If $M$, $m_W$, $m_H$ and $g_R^2(M^{-1})$ have 
been measured as in~\cite{bdsim1,bdsim2},
then 
\be
g^2_\rmi{$\msbar$}(m_W)\approx 
g_R^2(M^{-1})\biggl(1 + \fr12 \frac{m_W-M}{m_W} \biggr) 
+ \frac{|C(m_H/m_W)|}{158} g_R^4(M^{-1}).
\la{finres}
\ee
This relation constitutes our main result. 

\section{Implications for dimensional reduction}

As we recall from Sec.~1, 
in the finite temperature case
one is mainly interested in 
comparing results for non-perturbative quantities between
the 4d and 3d approaches. 
One of the non-perturbative quantities addressed is the critical
Higgs mass of the electroweak phase diagram. The phase diagram 
contains a line of first order phase transitions, 
which ends at a certain Higgs mass 
$m_{H,c}$~\cite{isthere},\cite{karschnpr}-\cite{bdsim3}.
A comparison of the 
4d and 3d results for $m_{H,c}$ yields perfect agreement at the 
percent level~\cite{bdsim2}. This comparison was possible
even before the relation of $g_R^2$ and $g_\rmi{$\msbar$}^2$ 
was known, since it turns out that $m_{H,c}$
depends only very weakly on $g_\rmi{$\msbar$}^2$~\cite{latt98}
(see also \fig\ref{fig:res})\footnote{In terms of 
powercounting, the dependence of $m_{H,c}$ on $g$ is of 
relative order ${\cal O}(g)$, 
so that the dependence on $g_\rmi{$\msbar$}^2 - g_R^2\sim {\cal O}(g^4)$ 
is only of relative order ${\cal O}(g^3)$.}.

Here we perform a comparison for another quantity, the 
critical temperature $T_c$. This comparison is in principle 
less powerful than that for $m_{H,c}$, since both the leading
($T_c \sim m_H/g$) and next-to-leading ($\delta T_c \sim m_H/(4\pi)$)
contributions are perturbative~\cite{pa} and thus definitely agree 
within the 4d and 3d theories. However, starting from the 
next-to-next-to-leading order ($\delta T_c \sim g m_H/(4 \pi)^2$), 
$T_c$ is non-perturbative~\cite{pa}.
(These statements can be easily understood from \eqs\nr{m32},\nr{msrel}.
Apart from perturbative terms within the 3d theory in \eq\nr{contaction},
the critical point is where $m_3^2 = c g_3^4 = c (g^2T)^2$, 
where $c$ is a non-perturbative coefficient which can 
only be determined numerically.)
Since $T_c$ can be measured with good accuracy with lattice 
simulations, a comparison becomes meaningful\footnote{
A similar comparison, but for a two scalar theory 
with very weak couplings, was effectively carried out in~\cite{jl}. 
However, due to the smallness of the couplings, that comparison was not 
sensitive to the non-perturbative terms in $T_c$.}. Due to the 
strong dependence on $g$, $T_c \sim m_H/g$, the relation of the
$g_R^2$ determined with 4d simulations and the $g^2_\rmi{$\msbar$}$
used in the 3d theory, is necessary for such a comparison: 
it is only sensible to discuss a non-perturbative
effect of relative order ${\cal O}(g^2)$,
once a perturbative ambiguity of the same relative order has been removed
by the computation carried out in this paper.

\begin{figure}[t]

\vspace*{-2cm}

\centerline{\epsfxsize=12cm\hspace*{-1cm}\epsfbox{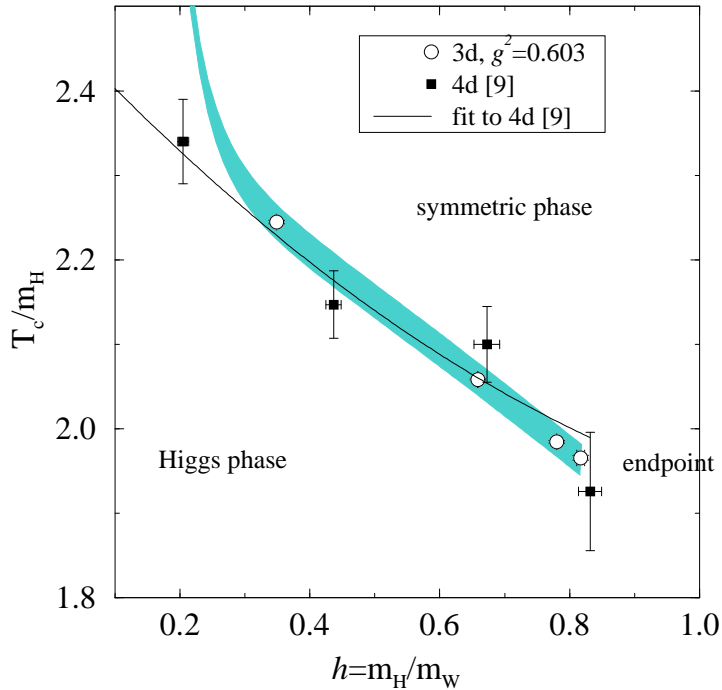}}

\vspace*{-5cm}

\caption[a]{A comparison of the phase diagrams
obtained with 4d and 3d simulations. The 4d results
are from~\cite{bdsim2}, and the 3d results 
from~\cite{nonpert,isthere,latt98}. 
The gray band has been obtained by subtracting
from the 3d lattice values the 3d perturbative values, fitting a 
curve to the difference, adding the perturbative curve
(which is very accurate within 3d at small $h$), 
and then converting to 4d physical units according to the formulas
in~\cite{generic}. The coupling was varied
in the interval $g^2_\rmi{$\msbar$}(m_W)=0.603\pm0.012$, the smallest 
$g^2_\rmi{$\msbar$}$ corresponding to the upper edge of the band.
The discrepancy at small $h$ is mainly due to the fact that the relation 
of $m_H,m_W$ to $m^2(\bmu), \lambda(\bmu)$
employed here and taken from~\cite{generic}, 
starts to break down as one approaches 
the Coleman-Weinberg limit $h\sim 0.13$, 
but also 
to the fact that the 3d theory itself is less accurate at small $h$ 
since the high-temperature expansion is not applicable in the case of 
an extremely strong transition.}  
\la{fig:res}
\end{figure}

With 4d lattice simulations, the continuum 
extrapolation for $T_c$ has been determined for 
$m_H = 34$ GeV in~\cite{bdsim1}, and for some other values, 
in particular $m_H = m_{H,c}$, in~\cite{bdsim2}.
The corresponding gauge coupling was measured to be
$g_R^2(M^{-1}) = 0.585(10)$ at $m_H=34$ GeV and, 
we interpret, consistent with this 
at $m_H = m_{H,c}$~\cite{bdsim2}. The mass parameter $M$ 
was observed to be close to the physical W mass $m_W$, 
but from Tables 4,5 in~\cite{bdsim1}, we observe that
there is a small discrepancy which, giving more weight 
to the lattices closest to the continuum limit ($L_t=4,5$), 
we estimate as $(m_W-M)/m_W\approx 0.035(25)$.
We now see from \eq\nr{finres} that 
this leads to $g_\rmi{$\msbar$}^2(m_W)\approx 0.603(12)$.
With this value, we can convert the 
3d results~\cite{nonpert,isthere,latt98} 
to be comparable with the 4d results, as explained in~\cite{4d3d}. 
The result is shown in \fig\ref{fig:res}.

We conclude that the phase transition line $T_c(m_H)$, 
and in particular
$m_{H,c}$, $T_c(m_{H,c})$, are in perfect agreement within statistical
errors at least for $h\gsim 0.3$ ($m_H\gsim 25$ GeV), 
whether determined directly from 4d or from a 3d effective theory.
For smaller~$h$, the vacuum renormalization employed~\cite{generic} 
in relating $m_H,m_W$ to $\msbar$ scheme breaks down, but even there 
dimensional reduction and the 3d theory themselves
are not completely off
in spite of the very strong transition, as we have checked 
separately by a vacuum renormalization which is accurate at the
Coleman-Weinberg point $h\approx 0.13$:
we get there $T_c/m_H\approx 2.5$.
Thus, dimensional reduction seems indeed as effective as 
one would have expected from the analytical estimates. 

\section{Conclusions}

In this paper, we have carried out a perturbative 1-loop computation
in the broken phase of the 4d SU(2)+Higgs theory, which establishes
a relation between the ``renormalized gauge coupling'' determined
by measuring the static potential with lattice Monte Carlo 
simulations, and the conventional $\msbar$ scheme gauge 
coupling $g^2(\bmu)$. With existing results for the mass 
parameter $m^2(\bmu)$ and scalar self-coupling $\lambda(\bmu)$, 
this relation completes what is needed for a reliable comparison
of 4d simulation results, and analytical computations 
where the $\msbar$ scheme is a natural starting point. As an important
application, we have considered the accuracy of 
finite temperature dimensional reduction, which is used, e.g., in 
the context of  
3d effective theory studies of the electroweak phase transition. 

The conclusion we find for dimensional reduction and the resulting
3d effective field theory is that, 
even for the non-perturbative characteristics of the electroweak phase
transition, the numerical accuracy is consistent with what it
was analytically estimated to be, i.e., at the 
percent level. This is quite good since, from the point of view 
of electroweak baryogenesis, studying the electroweak phase transition
with 3d effective theories and lattice simulations continues to be 
phenomenologically interesting in many extensions of the Standard Model, 
such as the MSSM~\cite{mssm2}.
 
\section*{Acknowledgements}

I thank Z. Fodor for providing me with the numerical values 
for the 4d datapoints in \fig\ref{fig:res}, and M. Shaposhnikov 
for useful discussions. This work 
was partly supported by the TMR network {\em Finite Temperature Phase
Transitions in Particle Physics}, EU contract no.\ FMRX-CT97-0122.

\section*{Note added}

I thank Y. Schr\"oder for pointing out that \eq\nr{res1}
agrees, for $N=2$, with the general result obtained for 
SU($N$)+Higgs in~\cite{ys}.

Very recently, a preprint appeared~\cite{cfhp} where the
same problem is addressed as in Sec.~\ref{sec:comp} of
the present paper. 

\appendix
\renewcommand{\thesection}{Appendix} 
 
\section{}
   
In this Appendix, we show the results for the individual graphs
in \fig\ref{graphs}. The results are given as contributions to 
the coefficient of $t$ in $\ln W(r,t)$, \eq\nr{Wrt}, and 
they are thus contributions to $-V(r)$ according to \eq\nr{potdef}.
We leave out a common factor 
\be
g^4 C_F C_A \int \frac{d^{d-1} p}{(2\pi)^{d-1}} 
\frac{e^{i\vec{p}\cdot\vec{r}}}{{p}^2 + m_W^2}
\ee
from all the non-Abelian terms (i.e., those proportional 
to $C_FC_A$). The Abelian terms (i.e., those proportional 
to $C_F^2$) only contribute to the exponentiation of
the leading order result in \eq\nr{Vtree}: they always
come with the coefficient $\fr12 t^2$ in $W(r,t)$
(see~\cite{wf,de} for a more precise discussion). Note that 
the explicit counterterm contribution, graph (g) in \fig\ref{graphs}, 
comes from the bare combination 
\be
g_B A_B = 
\biggl(
1- \frac{g^2 \mu^{-2\epsilon}}{(4\pi)^2\epsilon} \frac{3+\xi}{4} C_A
\biggr)g A,
\ee
where the numerical value is given for $d=4-2\epsilon$.

The contributions of the different graphs are:
\ba
& & \mbox{(a)}+\left.\mbox{(b)}\right|_\rmi{Abel.\ part} 
=  \mbox{exponentiation of the $r$-dependent term of \eq\nr{Vtree}}, \\
& & \left.\mbox{(b)}\right|_\rmi{non-Abel.\ part} = 
 (p^2 + m_W^2) \frac{d B_0(p^2;m_W^2,m_W^2)}{d m_W^2}  \nn 
& & \hspace*{1cm} + \frac{p^2 + m_W^2}{8(d-1) p^2 m_W^4}
    \biggl\{ - B_0(p^2,\xi m_W^2,\xi m_W^2) p^2 (p^2+4\xi m_W^2) \nn
& & \hspace*{1cm} + 2 B_0(p^2; m_W^2,\xi m_W^2)
    \Bigl[ p^4 + 2 p^2 m_W^2 (3+\xi-2d) + m_W^4 (\xi-1)^2
    \Bigr] \nn
& & \hspace*{1cm} - B_0(p^2;m_W^2,m_W^2) p^2 
    \Bigl[ p^2+4(3-2d) m_W^2\Bigr] \nn
& & \hspace*{1cm} + 2 m_W^2 (\xi-1) 
    \Bigl[ A_0(\xi m_W^2) - A_0(m_W^2) \Bigr]\biggr\}, \\
& & \mbox{(c)} + \left.\mbox{(d)}\right|_\rmi{Abel.\ part} =   
\mbox{exponentiation containing the cross-term 
from \eq\nr{Vtree},\hspace*{1.2cm}} \\
& & \left.\mbox{(d)}\right|_\rmi{non-Abel.\ part}  = 
     2 B_0(0;m_W^2,m_W^2) + \frac{1}{m_W^2}\Bigl[ 
     A_0(\xi m_W^2) -A_0(m_W^2) \Bigr], \\
& & \mbox{(e)} = 
\frac{1}{4(d-1) m_W^4 p^2} \biggl\{
  B_0(p^2,\xi m_W^2,\xi m_W^2)p^4 
      \left( {p^2} + 4\xi {m_W^2}\right) \nn
& & \hspace*{1cm} -
  2 B_0(p^2; m_W^2,\xi m_W^2)
  (p^2+ {m_W^2}) 
      \Bigl[ {p^4} + 2{p^2} {m_W^2}
         \left( 3 +\xi - 2d \right) + 
     {m_W^4} {{\left(\xi -1\right) }^2} 
         \Bigr] \nn
& & \hspace*{1cm} + B_0(p^2;m_W^2,m_W^2) p^2 
        \Bigl[ {p^4} + 
          2\left( 7- 4d \right) p^2 {m_W^2} + 
          8\left( 2- d \right) 
           {m_W^4}  \Bigr] \nn
& & \hspace*{1cm}  + 2 m_W^2 \Bigl[A_0(m_W^2) -  A_0(\xi m_W^2)\Bigr]  
      \left[ {p^2}\left( 4d -6 + \xi \right) + 
      {m_W^2}\left( \xi-1 \right)\right] \biggr\}, \\
& & \mbox{(f)} = \frac{-\Pi_W(p^2)}{C_A (p^2+m_W^2)} = 
\mbox{Eq.~\nr{res1}} - 
\left.\mbox{(b)}\right|_\rmi{non-Abel.\ part} - 
\left.\mbox{(d)}\right|_\rmi{non-Abel.\ part} - 
\mbox{(e)} \nn
& & \hspace*{1cm} -\,\Bigl\{\mbox{all $1/\epsilon$-divergences}\Bigr\}, \\
& & \mbox{(g)} =  -\frac{3+\xi}{2}
\frac{\mu^{-2\epsilon}}{(4\pi)^2\epsilon}.
\ea


\begin{thebibliography}{99}

\bibitem{rs}
V.A. Rubakov and M.E. Shaposhnikov,
Usp.\ Fiz.\ Nauk 166 (1996) 493 [hep-ph/9603208]. 

\bibitem{dr}
P.~Ginsparg,
Nucl.\ Phys.\ B 170 (1980) 388;
T. Appelquist and R. Pisarski,
Phys.\ Rev.\ D 23 (1981) 2305.

\bibitem{generic} K. Kajantie, M. Laine, K. Rummukainen and M. Shaposhnikov,
Nucl.\ Phys.\ B 458 (1996) 90 [hep-ph/9508379];
Phys.\ Lett.\ B 423 (1998) 137 [hep-ph/9710538].

\bibitem{an}
E. Braaten and A. Nieto,
Phys.\ Rev.\ D 53 (1996) 3421 [hep-ph/9510408].

\bibitem{jp}
A.~Jakov\'ac and A.~Patk\'os, 
Nucl.\ Phys.\ B 494 (1997) 54 [hep-ph/9609364].

\bibitem{mssm}
M. Laine, Nucl.\ Phys.\ B 481 (1996) 43 [hep-ph/9605283];
J.M. Cline and K. Kai\-nu\-lai\-nen,
Nucl.\ Phys.\ B 482 (1996) 73 [hep-ph/9605235];
Nucl.\ Phys.\ B 510 (1997) 88 [hep-ph/9705201];
M. Losada,
Phys.\ Rev.\ D 56 (1997) 2893 [hep-ph/9605266];
G.R. Farrar and M. Losada,
Phys.\ Lett.\ B 406 (1997) 60 [hep-ph/9612346].

\bibitem{nonpert}
K. Kajantie, M. Laine, K. Rummukainen and M. Shaposhnikov,
Nucl. Phys. B 466 (1996) 189 [hep-lat/9510020];
Nucl. Phys. B 493 (1997) 413 [hep-lat/9612006].

\bibitem{bdsim1} 
F. Csikor, Z. Fodor, J. Hein, A. Jaster and I. Montvay,
Nucl.\ Phys.\ B 474 (1996) 421 [hep-lat/9601016].

\bibitem{bdsim2}
F. Csikor, Z. Fodor and J. Heitger, 
Phys. Rev. Lett. 82 (1999) 21 [hep-ph/9809291].

\bibitem{isthere}
K. Kajantie, M. Laine, K. Rummukainen and M. Shaposhnikov,
Phys.\ Rev.\ Lett.\ 77 (1996) 2887 [hep-ph/9605288];
K. Rummukainen, M. Tsypin, K. Kajantie, M. Laine and M. Shaposhnikov,
Nucl.\ Phys.\ B 532 (1998) 283 [hep-lat/9805013].

\bibitem{4d3d}
M. Laine, 
Phys.\ Lett.\ B 385 (1996) 249 [hep-lat/9604011].

\bibitem{bdsim0} 
Z. Fodor, J. Hein, K. Jansen, A. Jaster and I. Montvay,
Nucl.\ Phys.\ B 439 (1995) 147 [hep-lat/9409017].

\bibitem{hh}
A. Hasenfratz and P. Hasenfratz, 
Phys. Lett. B 93 (1980) 165;
R. Dashen and D.J. Gross, 
Phys. Rev. D 23 (1981) 2340.

\bibitem{wf}
W. Fischler, 
Nucl.\ Phys.\ B 129 (1977) 157.

\bibitem{de}
M. Peter, 
Phys.\ Rev.\ Lett.\ 78 (1997) 602  [hep-ph/9610209];
Nucl.\ Phys.\ B 501 (1997) 471 [hep-ph/9702245];
Y. Schr\"oder, 
Phys.\ Lett.\ B 447 (1999) 321 [hep-ph/9812205].

\bibitem{karschnpr} 
F. Karsch, T. Neuhaus, A. Patk\'os and J. Rank,
Nucl. Phys. B (Proc. Suppl.) 53 (1997) 623
[hep-lat/9608087].

\bibitem{gurtler} 
M. G\"urtler, E.-M. Ilgenfritz and A. Schiller, 
Phys. Rev. D 56 (1997) 3888 [hep-lat/9704013];
E.-M. Ilgenfritz, A. Schiller and C. Strecha, 
hep-lat/9807023. 

\bibitem{ptw}
O. Philipsen, M. Teper and H. Wittig, 
Nucl.\ Phys.\ B 528 (1998) 379 [hep-lat/9709145].

\bibitem{aoki} 
Y. Aoki, 
Phys.\ Rev.\ D 56 (1997) 3860 [hep-lat/9612023];
Y. Aoki, F. Csikor, Z. Fodor and A. Ukawa, 
hep-lat/9901021.

\bibitem{bdsim3}
F. Csikor, Z. Fodor and J. Heitger,
Phys. Lett. B 441 (1998) 354 [hep-lat/9807021].

\bibitem{latt98}
M. Laine and K. Rummukainen, 
to appear in the Proceedings of {\sl Lattice '98},
July 1998, Boulder, Colorado [hep-lat/9809045].

\bibitem{pa} 
P. Arnold, 
Phys. Rev. D 46 (1992) 2628 [hep-ph/9204228].

\bibitem{jl}
K. Jansen and M. Laine,
Phys.\ Lett.\ B 435 (1998) 166 [hep-lat/9805024].

\bibitem{mssm2}
D. B\"odeker, P. John, M. Laine and M.G. Schmidt, 
Nucl.\ Phys.\ B 497 (1997) 387 [hep-ph/9612364];
M. Laine and K. Rummukainen, 
Phys.\ Rev.\ Lett.\ 80 (1998) 5259 [hep-ph/9804255];
Nucl.\ Phys.\ B 535 (1998) 423 [hep-lat/9804019];
Nucl.\ Phys.\ B 545 (1999) 141 [hep-ph/9811369];
M. Losada, 
Nucl.\ Phys.\ B 537 (1999) 3 
[hep-ph/9806519]; 
hep-ph/9905441.

\bibitem{ys}
Y. Schr\"oder, PhD thesis, DESY, Hamburg, 1999
(unpublished).

\bibitem{cfhp}
F. Csikor, Z. Fodor, P. Heged\"us and A. Pir\'oth, 
hep-ph/9906260. 

\end{thebibliography}
\end{document}